\documentclass[aps,twocolumn,epsf,floats,pre]{revtex4-1}
\usepackage{amsmath,amssymb}
\usepackage{graphicx,pstricks}
\usepackage{hyperref}
\usepackage{bm}
\include{epsf}

\def\(({\left(}
\def\)){\right)}                       
\def\[[{\left[}
\def\]]{\right]}

\newcommand{\<}{\langle}
\renewcommand{\>}{\rangle}
\newcommand{\beq}{\begin{equation}}
\newcommand{\eeq}{\end{equation}}
\newcommand{\bea}{\begin{eqnarray}}
\newcommand{\eea}{\end{eqnarray}}

\begin{document}
\title{Local equilibrium in bird flocks}

\author{Thierry Mora$^{1}$, Aleksandra M. Walczak$^{2}$, Lorenzo Del Castello$^{3,4}$, Francesco Ginelli$^{5}$, Stefania
  Melillo$^{3,4}$, Leonardo Parisi$^{6,4}$, Massimiliano
  Viale$^{3,4}$, Andrea Cavagna$^{4}$, Irene  Giardina$^{3,4}$
}

\affiliation{$^{1}$ Laboratoire de physique statistique,
    CNRS, UPMC and \'Ecole normale sup\'erieure, 24, rue Lhomond,
    Paris, France}
\affiliation{$^{2}$ Laboratoire de physique th\'eorique,
    CNRS, UPMC and \'Ecole normale sup\'erieure, 24, rue Lhomond,
    Paris, France}
\affiliation{$^{3}$ Dipartimento
    di Fisica, Universit\`a Sapienza, Rome, Italy}
\affiliation{$^{4}$ Istituto
    Sistemi Complessi, Consiglio Nazionale delle Ricerche, UOS
    Sapienza, Rome, Italy}
\affiliation{$^{5}$ SUPA, Institute
    for Complex Systems and Mathematical Biology, King’s College,
    University of Aberdeen, Aberdeen, UK}
\affiliation{$^{6}$ Dipartimento di Informatica, Universit\`a Sapienza, Rome, Italy}

\begin{abstract}
The correlated motion of flocks is an instance of global order emerging from local interactions. An essential difference with analogous ferromagnetic systems is that flocks are active: animals move relative to each other, dynamically rearranging their interaction network. The effect of this off-equilibrium element is well studied theoretically, but its impact on actual biological groups deserves more experimental attention. Here, we introduce a novel dynamical inference technique, based on the principle of maximum entropy, which accodomates network rearrangements and overcomes the problem of slow experimental sampling rates. We use this method to infer the strength and range of alignment forces from data of starling flocks. We find that local bird alignment happens on a much faster timescale than neighbour rearrangement. Accordingly, equilibrium inference, which assumes a fixed interaction network, gives results consistent with dynamical inference. We conclude that bird orientations are in a state of local {quasi-}equilibrium { over the interaction length scale}, providing firm ground for the applicability of statistical physics in certain active systems.

\end{abstract}

\maketitle



Animal groups moving in concert such as mammal herds, fish schools, 
and bird flocks show that in biology, just as in physics,
local coordination can result in large-scale order
\cite{Camazine2001,krause2002,sumpter2010}. 
However flocks differ from classical statistical physics 
in that their constituents are {active}: they constantly move by 
self-propulsion, pumping energy into the system and keeping it out of 
equilibrium \cite{Toner1998,Review, vicsek2012,marchetti2013}.
The key element is the rearrangement of the interaction network
due to the active motion of individuals relative to each other, continuously changing their neighbours.
Theoretical studies show that network rearrangement has major consequences,
which include enhancing collective order, reducing from 3 to 2  the lower critical dimension, 
and affecting the critical exponents \cite{Toner1995,Toner1998}. 

However, the importance of activity must be assessed with respect to
the relevant time scales of the system.
The impact of network rearrangement depends on the interplay between
its characteristic time scale, $\tau_{\rm network}$, defined as the average 
time it takes an individual to change its interaction neighbours,
and the local relaxation time scale, $\tau_{\rm relax}$, defined as the 
time needed to relax locally the order parameter if the interaction network were fixed.
If $\tau_{\rm network}\leq \tau_{\rm relax}$, the interaction network rearranges at least as fast as the order 
parameter relaxes, and the system remains far from equilibrium. If on
the other hand $\tau_{\rm relax}\ll  \tau_{\rm network}$, the
relaxation of the order parameter is adiabatic, closely following the
network as it slowly evolves. 
{In this case, even though the system behaves in an
out-of-equilibrium manner on the longest scales, it locally obeys a condition of equilibrium,  
and we expect some of the tools of equilibrium statistical physics to be applicable.} 

Here, we explicitly address the impact of network activity by developing a new inference method based on the exact integration 
of maximum-entropy dynamical equations, thus accounting for the reshuffling of the network.  We apply the method to data 
of starling flocks of up to 600 individuals \cite{Ballerini2008a,Cavagna2008b,Cavagna2008a, Attanasi2015} 
(see Materials and Methods and Table S1 for data summary), inferring the relevant 
parameters of the interactions between individuals. We find that the alignment relaxation time, 
$\tau_{\rm relax}$, is more than one order of magnitude shorter than the network rearrangement time, 
$\tau_{\rm network}$.  Consistently, we show that the parameters learned from the dynamics are consistent
with those obtained by an equilibrium-like inference, which assumes a fixed network \cite{Bialek2012}. 
Our results suggest that natural flocks are in a state of local
quasi-equilibrium
{ over the interaction length scale,}
meaning that the relatively slow 
rearrangement of the local interaction network does not affect the ordering dynamics {up to certain scales}.

\begin{figure*}
\includegraphics[width=\linewidth]{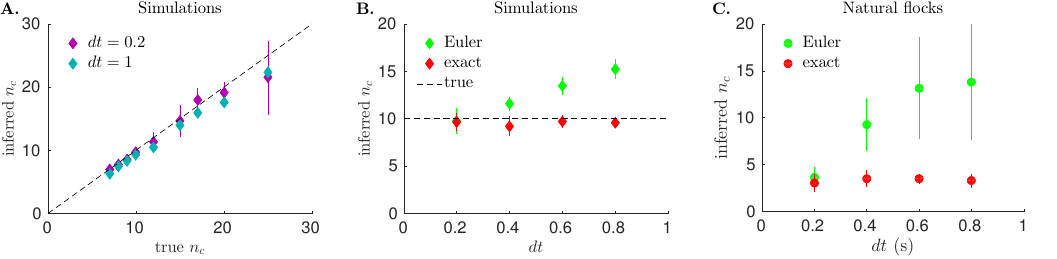}
\caption{Performance of the inference methods on the predicted interaction range $n_c$.
{\bf A.} Inferred versus real $n_c$ obtained by applying our new inference method to simulated data generated with Eq.~\ref{model} at various interaction ranges. The method performs well for different values of the sampling rate $dt$.
{\bf B.} Dependence of the inferred $n_c$ on the sampling time $dt$. On simulated data with $n_c=10$ (dashed line), the inference method based on exact integration (red points) performs well regardless of the sampling time $dt$. By contrast, the inference method based on Euler's integration method (green points) overestimates the true interaction range at large $dt$.
{\bf C.} A similar trend is observed when we apply the two inference
procedures to real flocking data, as illustrated here on one flocking
event. Note that in this case the true value is not
known. { Error bars represent standard errors over time frames.}
\label{testinference}
}
\end{figure*}

To compare the relevant time scales of the ordering process
in flocks, we first need to
learn the dynamical rules of their behaviour.
Learning these rules usually relies on inferring the 
parameter of a chosen model directly from the data,
as has been recently done in surf scoters \cite{Lukeman:2010p9518}, prawns \cite{mann2012} and fish
schools \cite{Katz2011a,Herbert-Read2011,Gautrais2012,strandburg2013,Rosenthal2015}. 
Although in these studies  the local rules of interaction were often learned using small groups, 
in some cases they could also be used to predict large-group behavior \cite{Gautrais2012,Rosenthal2015}. 
Here, instead of assuming a model {\it a priori}, we apply the principle of maximum entropy to the trajectories 
of all birds in the group \cite{Cavagna2013}.
We look for a distribution of the stochastic process that is as random as possible, 
while agreeing with the data on a key set of experimental observables. 

In a flock of size $N$, we call $\vec s_{i}(t)$ the three-dimensional flight orientation of bird $i$ at time $t$. 
The maximum entropy distribution over possible flock trajectories that is consistent with the correlation 
functions $\<\vec s_i(t)\cdot \vec s_j(t)\>$, as well as their derivatives $\<d\vec s_i(t)/dt\cdot \vec s_j(t)\>$, 
can be exactly mapped, in the limit of strong polarization $P\equiv (1/N)\Vert \sum_i \vec s_i\Vert \approx 1$, 
onto the following stochastic differential equation (see SI and Ref.~\cite{Cavagna2013}):
\beq
\label{model}
\frac{d\vec s_i}{dt}=\left(\sum_{j} J_{ij} \vec s_j+\vec\xi_i\right)_\perp,
\eeq
where $\vec\xi_i$ is a random white noise, and
where the projection $\vec x_\perp\equiv \vec x - \vec s_i(\vec x\cdot \vec s_i)$ onto the plane perpendicular to $\vec s_i$ ensures that $\vec s_i$ remains of norm 1. Equation \eqref{model} can be viewed as a generalization of the Vicsek model \cite{Vicsek1995}:
each bird modifies its flight direction according to a weighted average of the directions of its neighbours. The interaction matrix $J_{ij}$ encodes 
how much bird $i$ is influenced by ({\em i.e.} interacts with) bird $j$. Given the experimentally measured correlation functions, entropy maximization yields equations that fix the values of  the noise amplitude and the interaction matrix $J_{ij}$. 
This matrix has too many parameters to be 
reliably determined from the data, but we can reduce its complexity by parametrising it. It was shown in 
\cite{Cavagna2015a} that the interaction decays exponentially with the topological distance 
$k_{ij}$ between birds,
\beq
J_{ij}=J \exp(-k_{ij}/n_c) \ , 
\label{bongo}
\eeq
where $k_{ij}$ denotes the (time-dependent) rank of bird $j$ among the neighbours of bird $i$ 
ranked by distance. This interaction matrix has just two parameters: $n_c$ is the 
topological interaction range, while $J$ is the overall strength of the interaction.
The noise is uncorrelated among birds and of uniform magnitude $T$, 
by analogy with physical temperature: $\<\vec\xi_i(t)\cdot\vec\xi_j(t')\>=2\, d\, T\,\delta_{ij} \delta(t-t')$, where $d$
is the space dimension ($d=3$ in the following).

In principle, to learn the parameters of Eq.~\ref{model} one needs actual 
continuous-time derivatives. In practice, we only have  configurations separated by the
finite experimental sampling time $dt$. A common solution is to use 
Euler's approximation:
\beq\label{linapprox}
\vec s_i(t+dt)\approx \vec s_i(t)+dt\,\sum_{j} J_{ij} \vec {s_j}_\perp+\sqrt{2T dt}\, \vec{\eta_i}_\perp,
\eeq
where $\vec \eta_i$ is a normally distributed vector of variance 1 in each direction.
The conditional likelihood of the data given the model, $P[\{\vec s_i(t+dt)\}|\{\vec s_i(t)\}]$, can be written in Gaussian 
form after expanding Eq.~(\ref{linapprox}) in the spin-wave approximation  (see Materials and 
Methods). Maximising this likelihood yields  values for the alignment parameters $n_c$, $J$ and $T$ (see 
Ref.~\cite{Cavagna2013} and SI).  

Euler's approximation is used by virtually all methods that try to fit
a dynamical equation to a discrete time  series
\cite{Katz2011a,Herbert-Read2011,Gautrais2012}.
However, it is inappropriate
when the experimental sampling time, $dt$, is larger than the
intrinsic relaxation timescale, $\tau_{\rm relax}$. In this case
information spreads between subsequent frames beyond the directly interacting neighbours and
Euler's approximation {overestimates} the range of the interaction, as
we shall see below. 
To overcome this issue,
we rewrite Eq.~\ref{model} by formally subtracting $\sum_{l}J_{il} \vec {s_i}_{\perp}=0$ from it:
\beq
\label{model2}
\frac{d\vec {\bf s}}{dt}=-J{\bm \Lambda} \vec {\bf s}_\perp+\vec{\bm
    \xi}_\perp.
\eeq
Bold symbols denote vectors and matrices over bird indices; the
matrix $\Lambda_{ij}\equiv\delta_{ij}\sum_{l}n_{il}-n_{ij}$, 
where $n_{ij} = e^{-k_{ij}/n_c}$ is the connectivity
matrix \eqref{bongo}. ${\bm \Lambda}$ is
analogous to a Laplacian defined on a lattice, and obeys the sum rule: $\sum_j\Lambda_{ij}=0$.
In the spin-wave approximation, where all orientations $\vec
s_i$ point in almost the same direction, this relation ensures
that ${\bm \Lambda} \vec {\bf s}$ has almost no contribution along the
common direction of flight, implying
$({\bm \Lambda} \vec {\bf s})_{\perp}\approx {\bm \Lambda} \vec {\bf
  s}$ (see Materials and Methods and SI).
Equation \ref{model2} is now linear and it can be  integrated exactly:
\beq\label{exactint}
\vec{\bf s}(t+dt)=e^{-J{\bm \Lambda}dt}\vec{\bf s}(t) + \int_0^{dt} du\,e^{-J{\bm \Lambda}(dt-u)}\vec{\bm \xi}_\perp(t+u) \ .
\eeq
This result assumes a constant $J_{ij}$ in the interval $dt$, which is a good approximation if $dt\ll \tau_{\rm network}$. 
Fortunately, this same condition is necessary for the very possibility 
to collect data: tracking requires to follow each individual across time, 
which is only possible if individuals do not significantly change
their neighbourhood between consecutive frames. 
The integrated noise in the right-hand side of \eqref{exactint}  is Gaussian, of mean zero and 
covariance $4T \int_0^{dt} du\,e^{-J{\bm \Lambda}u} e^{-J{\bm \Lambda}^\dagger u}$. 
Using the exact solution \eqref{exactint} we can write an explicit expression for the (Gaussian) conditional likelihood 
$P[\{\vec s_i(t+dt)\}|\{\vec s_i(t)\}]$, which can then be maximised over the parameters of the model  
(see Materials and Methods). 

We first tested our dynamical inference method on synthetic data simulated
using the model of Eq.~\ref{model}, with $\tau_\mathrm{relax}\approx
0.7$, for various values of the interaction range 
$n_c$ (see Materials and Methods). We infer the parameters of the model using either Euler's rule or the result of exact integration, 
for different values of the sampling time ranging from $dt=0.2$ to
$dt=0.8$. The method based on exact integration predicts the interaction range $n_c$ well, regardless of $dt$ (Fig.~\ref{testinference}A and B),
while the method based on Euler's approximation largely overestimates
$n_c$ at large $dt$ (Fig.~\ref{testinference}B). We can now apply our dynamical inference to real flocks and learn the model parameters.
First, we used data of natural flocks to check the effect of
changing the sampling time $dt$, from
the real sampling time of our setup, $dt=0.2$\,s (see Materials and
Methods), to $0.8$\,s.
Although we cannot 
compare the inferred value of $n_c$ to the ground truth as in simulations, we observe a similar trend as a function of $dt$ 
(Fig.~\ref{testinference}C), with the exact integration and
Euler's approximation methods agreeing only at small $dt$.
This suggests that the sampling time 
of $0.2$\,s is of the same order as the orientation relaxation time
$\tau_{\rm relax}$, as we will confirm below. It also 
indicates that the inference method based on exact integration is extracting the parameters of alignment reliably.

Using the model parameters learned from the data, we 
evaluate the two time scales of interest for activity, namely relaxation of the orientations 
and network rearrangement.
We estimated the network rearrangement time $\tau_{\rm network}$
experimentally for each flocking event as the characteristic decay
time of its autocorrelation function $C_{\rm
  network}(t)=\sum_{ij}n_{ij}(t_0)n_{ij}(t_0+t)$, by fitting $C_{\rm
  network}(t)\approx C_0\exp(-t/\tau_{\rm network})$ (Fig.~\ref{overlap}). 

Working out the time scale of relaxation is more subtle.
The relevant quantity is the product of the interaction strength $J$,
which has units of inverse time,
by the dimensionless connectivity matrix, $\bm \Lambda$, as can be seen from Eq.~\eqref{model2}. 
Since there are $n_c$ neighbours acting on each individual, the total alignment force is of order $Jn_c$, 
suggesting that the characteristic time scale of relaxation of 
the orientations is $\tau_{\rm relax} \sim (Jn_c)^{-1}$. 
This result, however, seems at odds with the well-known fact that
systems with spontaneously broken
continuous symmetry - such as flocks - have correlation length and
relaxation time that diverge with the system size $L$ (Goldstone theorem \cite{Parisi}).
On the other hand, we do not expect the large-scale modes responsible for this divergence to affect the local
relaxation dynamics and its interplay with network reshuffling.
To clarify this issue we calculate the dynamical autocorrelation function of the fluctuations
of the order parameter, $C_\mathrm{relax}(t) = \< \delta \vec s_i(t_0)\cdot \delta\vec s_i(t_0+t)\>$,
where $\delta \vec s_i = \vec s_i - \<\vec s_i\>$. 
We consider a fixed lattice, because we need to gauge 
relaxation in absence of network rearrangements, resulting in the autocorrelation function
(see SI):
\beq
\label{bobmain}
C_\mathrm{relax}(t) = \int_{1/L}^{1/a} d^dk  \ \frac{e^{-J a^2 n_c \, k^2 t}}{J a^2 n_c \, k^2} \ ,
\eeq
where $a$ is the lattice spacing. The infrared divergence at small $k$, 
which correspond to large-scale modes, makes the integral divergent in the 
$L\to\infty$ limit for $d=2$ (Mermin-Wagner theorem \cite{Mermin}).
In $d=3$ the integral is finite, but 
the correlation function is a power law, so that the relaxation time
diverges with $L$ (Goldstone theorem).
{
The small $k$ modes in \eqref{bobmain} correspond to 
long wavelengths fluctuations spanning the entire flock, 
causing the local order parameter to relax slowly.
However, these long wavelength fluctuations do not contribute to the disordering of the local interaction network: if the 
wavelength of a fluctuation is much larger than the {\it interaction} range, all directions of motion in the interaction
neighbourhood fluctuate in unison, causing no change in the mutual positions of the 
birds. 
We conclude that the autocorrelation function that impacts on local network rearrangements
only includes contributions from wavelengths up to the local interaction range (let us call it $r_c$). This amounts to restricting 
the integral in \eqref{bobmain} to the modes $r_c^{-1}\leq k\leq a^{-1}$, thus eliminating the infrared divergent modes
$k\sim 1/L$. The resulting correlation function is exponentially decaying (see SI for the calculation of the integral), 
with finite relaxation time equal to $\tau_\mathrm{relax}=(J n_c)^{-1}$, consistent with our initial guess.
We note that, by considering wavelengths up to the interaction range, we are still dealing with
a coarse-grained field theory, as in most biological systems the scale
of interaction extends over tens of neighbours.}

\begin{figure}
\begin{center}
\includegraphics[width=\linewidth]{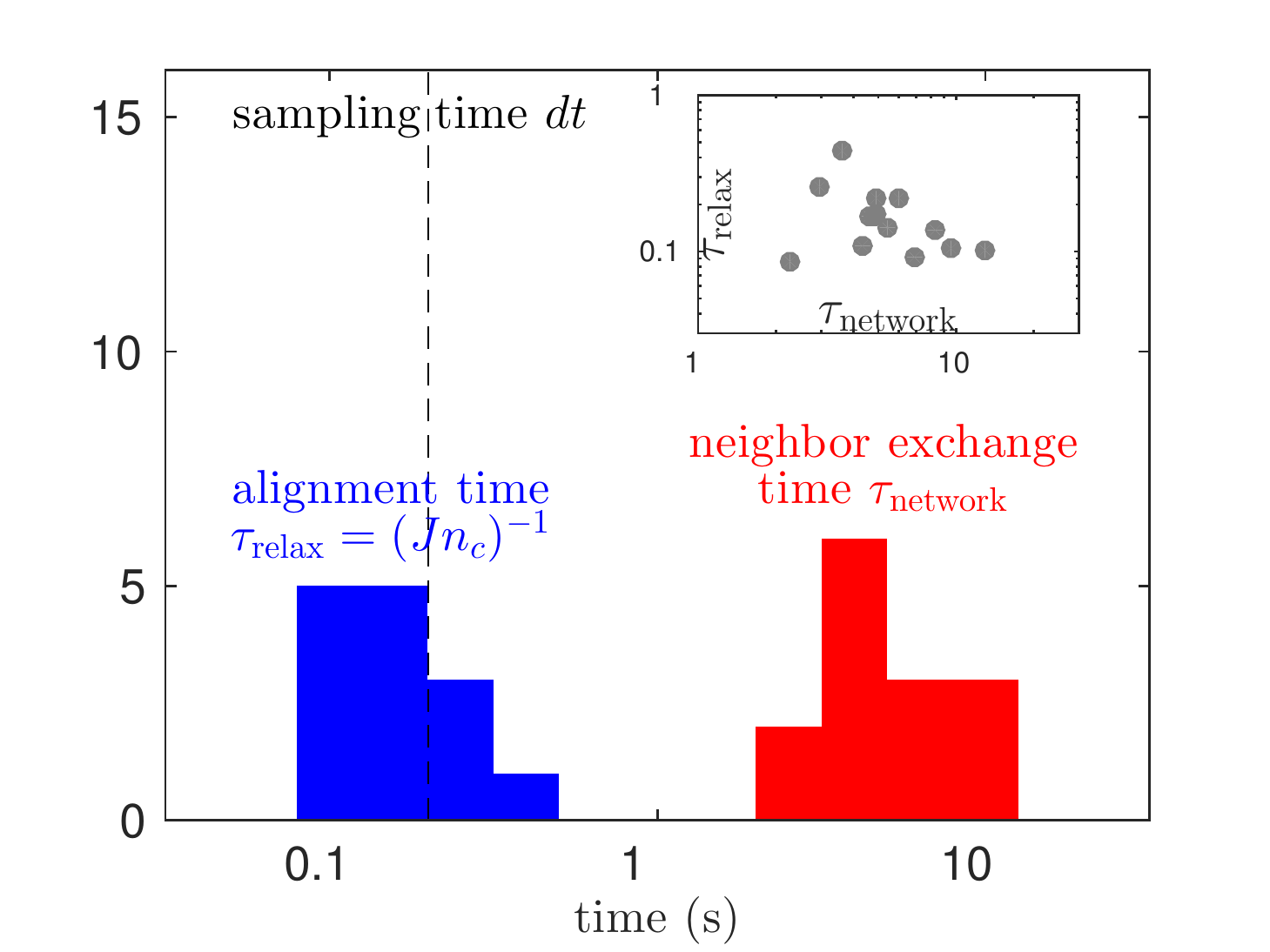}
\caption{Comparison between the two relevant time scales of active
  matter, as inferred in 14 natural flocks using our inference method
  based on exact integration. Histograms of the neighbour exchange
  time $\tau_{\rm network}$ versus the local alignment time $\tau_{\rm
    relax}=1/Jn_c$, show that the relaxation of orientations is much
  faster than the turnover of neighbours. Note that the experimental
  sampling time $dt=0.2$\,s (dashed line) is of the same order as the
  alignment time, justifying the use of exact
  integration. { Inset: the scatter plot of $\tau_{\rm
      relax}$ versus $\tau_{\rm network}$ shows no correlation between
    the two quantities.}
\label{timescales} }
\end{center}
\end{figure}

We can now proceed with the
comparison of $\tau_{\rm network}$ and
$\tau_{\rm relax}$. Results are 
summarised in Fig.~\ref{timescales}. The two time scales clearly separate, with
local relaxation almost two orders of magnitude faster than network reshuffling. 
This separation of time scales suggests that flocks are in a state of local equilibrium. The network of interactions
changes slowly enough for the dynamics of flight orientations to catch up before neighbours
reshuffle.  In other words, the orientation dynamics tracks network changes adiabatically. 
Note that this statement holds only locally,
at the scale of the interaction range, as both $\tau_{\rm network}$ and
$\tau_{\rm relax}$ are defined on that scale.

\begin{figure*}
\begin{center}
\includegraphics[width=\linewidth]{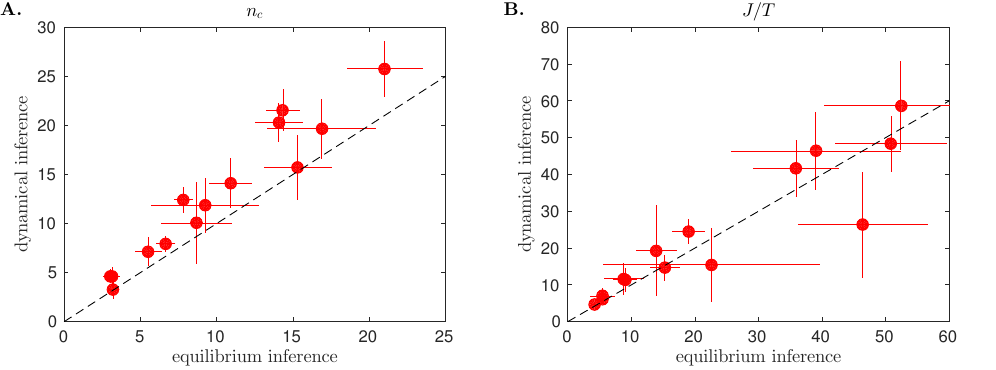}
\caption{Inference on natural flocks. For each of the 14 flocking
  events, the parameters of the model were inferred using either the
  dynamical inference method presented here, with $dt=0.2$\,s, or an
  equilibrium inference method as in \cite{Bialek2012}. {\bf A}. Both
  methods agree well on the predicted value of the alignment range
  $n_c$. {\bf B.} While the dynamical method infers the alignment
  strength $J$ and the noise amplitude $T$ separately, the equilibrium
  method only infers their ratio $J/T$, the value of which is
  consistent between the two methods. { Error bars represent
    standard errors over time frames.}
\label{results}}
\end{center}
\end{figure*}

Since flocks behave as if they were in local equilibrium, an equilibrium
inference procedure, which takes as input the local spatial correlation computed 
from a snapshot of the birds'  flight orientation \cite{Bialek2012}, should be consistent with the 
results of the dynamical inference.  To check this prediction, we recall the equilibrium-like inference method of \cite{Bialek2012}. For symmetric $J_{ij}$,
Eq.~\ref{model} is the Langevin equation derived from the Hamiltonian of the Heisenberg model
\beq
\mathcal{H}=-\frac{1}{2}\sum_{i,j}J_{ij}\vec s_i\cdot \vec s_j.
\eeq
When $J_{ij}$ varies slowly in time, the fluctuations of $\vec s_i$
are in quasi-equilibrium and distributed according to Boltzmann's law:
\beq\label{static}
P(\vec s_1,\ldots,\vec s_N)\sim \exp\left(-{\mathcal{H}}/{T}\right).
\eeq
We recognise the maximum entropy distribution consistent with the local correlation index 
$\sum_{ij}n_{ij}\<\vec s_i\vec s_j\>$ fitted in
Ref.~\cite{Bialek2012}.
In practice, the equilibrium inference consists in maximising the likelihood of Eq.~\ref{static} over its parameters $n_c$ and $J/T$ (see Materials and Methods and SI).
If the variations of $n_{ij}$ are slow compared to the dynamics of $\vec s_i$, $\tau_{\rm network}\gg \tau_{\rm relax}$, this
inference procedure should give an accurate estimate of the alignment parameters.
{
If however the two time scales are comparable, we expect the
equilibrium inference to overestimate the true $n_c$, as the frequent
exchange of neighbours results in an effective number of interaction
partners that is larger than the instantaneous one. We verified both
these expectations on simulated data, by showing that the equilibrium
inference is accurate for $\tau_{\rm network}\sim 100\tau_{\rm
  relax}$, but overestimates $n_c$ for $\tau_{\rm network}\sim \tau_{\rm relax}$ (see Fig.~\ref{testtimescales}).
}
{ When applied to empirical data, the
dynamical and equilibrium
inferences give consistent results, and predict the same interaction
range, $n_c$, and coupling-to-noise ratio, $J/T$ (Fig.~\ref{results})
Note that, while the dynamical inference provides the strength of the interaction, $J$, and the strength of the noise, $T$, 
separately, the equilibrium inference only gives the ratio $J/T$, which is the quantity to compare.}
To better appreciate
this result, recall that the two inference procedures are based on
independent pieces of information: the equilibrium inference uses instantaneous orientations, while the dynamical inference exploits
how these orientations change in time. 
Their agreement confirms that the alignment dynamics of flocks are in an
effective state of equilibrium { over the range $n_c$}.

{Theoretical studies of active matter indicate that out-of-equilibrium
effects induced by the rearrangement of the interaction network play a
major role in the ordering of the system  \cite{Toner1998,Review}. 
In this light, any attempt to understand the properties of active 
biological systems based on equilibrium approaches may seem inappropriate.
Does it mean that we should we always relinquish
the methods of equilibrium statistical mechanics when dealing with active systems?
Our results address this question by showing that bird flocks are 
in a state of local equilibrium, due to the rapid relaxation of orientations compared to the slow
rearrangement of the network,  { over the local scale of interaction}. As a consequence, an equilibrium
inference method, which assumes a fixed
interaction network, gives equivalent results to a full
dynamical treatment.

Equilibrium inference seems to be justified in this system, 
not only as a formal mathematical equivalence allowing for useful insights and predictions, but as a tool to extract 
{\it bona fide} biological parameters. The equilibrium approach is mathematically 
simpler and computationally less expensive than the dynamical one in the limit of strong polarisation,
making it easier to analyse larger groups. Although a dynamical
approach such as the one presented here is still necessary for extracting the precise relaxation timescale of the ordering mechanism, there may be more straightforward ways to evaluate its order of magnitude and get a quick assessment  of the local equilibrium hypothesis.
}

Our results do not mean that natural flocks {are in global
  equilibrium} and that {network rearrangements} play no role.
{ The interaction network, far from being fixed as if individuals were linked by springs \cite{Ferrante2013}, completely reshuffles on long time scales \cite{Cavagna2013a}.}
The directions of motion 
relax on a faster time scale than the network over the local scale of interaction, but the network {does} move on longer time scales, {and over larger length scales,}
with important consequences. To appreciate this point we must stress again the difference between
local, short-wavelength modes, which set the
balance between relaxation and network rearrangement, and long-wavelength 
modes, which govern the long time and long distance correlations. Capturing these large-scale
properties requires to describe the active fluid 
using a hydrodynamic approach \cite{Toner1998}. { Equilibrium inference works despite the existence of these large-scale modes because it only uses  information at the local scale of interaction, where relaxation is fast.}

The local equilibrium we have uncovered in natural flocks is not
merely the consequence of the high degree of polarisation of this
system. A high polarisation certainly implies slow network 
rearrangements, but it does not constrain the 
relaxation time,
which could be even slower, { as illustrated in our simulations (Fig.~\ref{testtimescales}}). Conversely, there 
may be unpolarised systems where local relaxation is faster than network rearrangement { -- a limit easily obtained theoretically by considering weakly interacting, slowly moving individuals. Midge swarms may be such an example: they are not polarized, poised below the ordering transition \cite{Attanasi2014a}, yet have been successfully analysed using standard equilibrium tools of critical phenomena \cite{Attanasi2014b}.}
In general, one must carefully quantify these two time scales to determine 
to what degree the tools of equilibrium
statistical mechanics may be applied to a given active system.

\section*{Materials and methods}

\subsection*{Flocking data.}
The three-dimensional trajectories of all birds were reconstructed
using imaging techniques. Stereoscopic experiments on natural flocks
of European starlings were performed in the field in Rome using three
high speed machine vision cameras shooting at 170 fps . The
stereoscopic video acquisitions were then processed using a novel
purpose-built three-dimensional tracking algorithm based on  a
recursive global optimization method \cite{Attanasi2015}. This
algorithm is extremely powerful, allowing for the reconstruction of
full 3D trajectories of all individuals in groups of several hundreds
individuals. We collected 3D data from 12 flocking events with sizes
ranging from 50 to 600 individuals, and lasting from 2s to 6s (for
details on the experiments and the dataset see Table S1 and \cite{Cavagna2008b,attanasi2014}).
To avoid interference from birds flapping, which occurs at frequency $\approx 10$\,Hz, we 
subsampled all the 3D sequences so that two snapshots are separated by $dt'=0.1$~s. 
The instantaneous flight orientations were estimated by
$
\vec s_i(t)=[{\vec r_i(t+dt')-\vec r_i(t)}]/{\Vert r_i(t+dt')-\vec r_i(t)\Vert}.
$
To avoid overlap between two subsequent evaluations of $\vec s_i(t)$, we used $dt=2dt'=0.2$~s. The lower sampling rates of Fig.~\ref{testinference}C, were obtained by taking $dt'=0.2$, $0.3$, and $0.4$~s.

\subsection*{Simulated data.}
Data were simulated in three dimensions with the continuous Vicsek model of Eq.~\ref{model} with the interaction matrix of Eq.~\ref{bongo}.
The positions $\vec r_i$ of individuals are updated
  according to $d\vec r_i/dt=v_0 \vec s_i$, with $v_0=1$.
The simulations were set in a $8\times 8\times 8$ box with periodic
boundary conditions, and $N=512$ birds, so that density is exactly
1. We set $\sqrt{2T}=0.15$ to obtain
a polarization $P\approx 0.99$ similar to natural flocks.
Eq.~\ref{model} was integrated using Euler's method with a simulation step $dt_{\rm sim}=0.01$ that is much smaller than any other time scale in the system. The interaction range $n_c$ varied from $7$ to $25$, and the interaction strength was picked so that $Jn_c=1.5$, hence $\tau_\mathrm{relax} = (Jn_c)^{-1} \sim 0.7$.
The flocks were first brought to a steady state before taking snapshots for analysis.

\subsection*{Spin-wave approximation.}
The polarization $P$ quantifies the level of order in the system. When $P\approx 1$, we can expand each $\vec s_i$ around the common direction of flight $\vec n\equiv (1/NP)\sum_i\vec s_i$. This expansion gives $\vec s_i = \vec \pi_i + \sqrt{1-\vec \pi_i^2} \vec n \approx  \vec \pi_i + (1-\vec \pi_i^2/2) \vec n$, with $\vec n\cdot \vec \pi_i=0$. At leading order in $\vec \pi_i\ll 1$, Eq.~\ref{model2} becomes
\beq\label{dynpi}
\frac{d\vec \pi_i}{dt}=-J \sum_{j} \Lambda_{ij} \vec\pi_j+\vec\xi_{i\perp},
\eeq
with $\<\vec\xi_{i\perp}(t)\vec\xi_{j\perp}(t')\>=4T\delta_{ij}\delta(t-t')$.
Similarly, the equilibrium distribution (Eq.~\ref{static}) can be expanded into
\beq\label{statpi}
P(\vec {\bm \pi})
=\frac{1}{Z}e^{-(J/T) \sum_{ij} \Lambda_{ij}\vec \pi_i\cdot \vec\pi_j}.
\eeq
Since this distribution is Gaussian, $Z$ can be calculated analytically and reads: $Z=({2\pi T}/{J})^{(N-1)}\prod_{\lambda_k>0} \lambda_k^{-1}$,
where $\lambda_k$ are the eigenvalues of the matrix $\Lambda_{ij}$.

\subsection*{Maximum likelihood Inference.}
The {\em equilibrium inference} is performed by maximising the likelihood of the data given by Eq.~\ref{statpi} over the parameters $n_c$ and $(J/T)$ (see SI for detailed formulas).

The {\em dynamical inference based on Euler's rule} is implemented by maximising the likelihood $P(\{\vec \pi_i(t+dt)\}|\{\vec \pi_i(t)\})$ calculated from Euler's formula (Eq.~\ref{linapprox}). This likelihood reads
\beq\label{Leuler}
{(4\pi Tdt)}^{-N}e^{-\frac{1}{4Tdt}\sum_i[\vec \pi_i(t+dt)-\vec \pi_i+Jdt \sum_{j} \Lambda_{ij} \vec\pi_j ]^2}.
\eeq

The {\em dynamical inference based on exact integration} uses
Eq.~\ref{exactint}, rewritten as $\vec{\bm \pi}(t+dt)=e^{-J{\bm \Lambda}dt}\vec{\bm \pi}(t) + \vec{\bm\epsilon}$,
where 
$\vec{\bm\epsilon}$ is a zero-mean Gaussian vector of covariance
$\<\vec{\bm \epsilon} \vec{\bm \epsilon}^{\dagger}\>= 4T \int_0^{dt} du\,e^{-J{\bm \Lambda}u}e^{-J{\bm \Lambda}^\dagger u}={\bf X}^{-1}$.
The conditional likelihood $P(\{\vec \pi_i(t+dt)\}|\{\vec \pi_i(t)\})$ now reads
\beq\label{Lexact}
\frac{\mathrm{det}({\bf X})}{(2\pi)^{N}}e^{-\frac{1}{2}[\vec{\bm \pi}(t+dt)-e^{-J{\bm \Lambda}dt}\vec{\bm \pi}(t)]^\dagger {\bf X}[\vec{\bm \pi}(t+dt)-e^{-J{\bm \Lambda}dt}\vec{\bm \pi}(t)]}.
\eeq
Depending on whether one uses Euler's or exact integration rules, Eq.~\ref{Leuler} or \ref{Lexact} is maximised over $J$, $T$ and $n_c$ (see SI for detailed formulas).

In all three inference procedures, the parameters are learned for each time $t$. Then the median and the associated standard error are calculated for each flocking event.

{
\subsection*{Data Availability.}
The data that support the plots within this paper and other findings of this study are available from the corresponding author upon request.
}

\bigskip

\subsection*{Acknowledgements.}
Work in Paris was supported European Research Council Starting Grant 306312. Work in Rome was supported by IIT-Seed Artswarm, European Research Council Starting Grant 257126, and US Air Force Office of Scientific Research Grant FA95501010250 (through the University of Maryland). F.G. acknowledges support from EU Marie Curie ITN grant n. 64256 (COSMOS) and Marie Curie CIG PCIG13-GA-2013-618399.

\bibliographystyle{pnas}

\appendix

\setcounter{figure}{0}
\makeatletter 
\renewcommand{\thefigure}{S\@arabic\c@figure}
\makeatother

\setcounter{table}{0}
\makeatletter 
\renewcommand{\thetable}{S\@arabic\c@table}
\makeatother

\section{Dynamical maximum entropy model}
Call $\vec s_i(t)$ the $d$-dimensional flight orientation of bird $i$
as a function of time, of unit norm $\Vert \vec s\Vert=1$. We look for
a probability disribution over whole flock trajectories,
$(\vec s_1(t),\ldots,\vec s_N(t))$, that has maximum entropy, but with the constraints that the correlation functions:
\beq
\<\vec s_i(t)\cdot \vec s_j(t)\>
\eeq
and
\beq
\left\<\frac{d\vec s_i(t)}{dt}\cdot \vec s_j(t) \right\>
\eeq
agree with the data.
After time discretization, these
constraints are equivalent to imposing the values of $\<\vec
s_i(t)\cdot \vec s_j(t)\>$ and $\<\vec s_i(t+dt)\cdot \vec s_j(t)\>$,
with $dt$ an infinitesimal increment. Using the technique of Lagrange
multipliers, one can show that the distribution over
trajectories then takes the form \cite{Jaynes1,Jaynes2}:
\beq
\begin{split}
P(\{\vec s_i(t)\})&=\frac{1}{\mathcal{Z}} \exp\left(\sum_{ij,t}
  J^{(1)}_{ij;t}\vec s_i(t) \cdot \vec s_j(t)\right.\\
&\left.+\sum_{ij,t}
  J^{(2)}_{ij;t}\vec s_i(t+dt) \cdot \vec s_j(t)\right)\prod_{i,t} \delta(\Vert \vec s_i(t)\Vert-1)
\end{split}
\eeq
where sums and products over $t$ run over a discrete set of times
separated by $dt$,  and where $\delta(\cdot)$ denotes the Dirac-delta function.

In \cite{Cavagna2013}, it was shown that, in the spin-wave
approximation, the stochastic process described by this probability
distribution is equivalent to a random walk:
\beq
\vec s_i(t) = \frac{\sum_j M_{ij;t} \vec s_j(t) +\vec\eta_i(t)}{\Vert\sum_j M_{ij;t} \vec s_j(t) +\vec\eta_i(t)\Vert},
\eeq
with $\eta_i(t)$ is a Gaussian variable of zero mean and covariance
$\<\eta_i(t)\cdot \eta_j(t')\>=d (A_t^{-1})_{ij}\delta_{t,t'}$. The
matrices $M_{ij;t}$ and $A_{ij;t}$ can be expressed in terms of the
matrices $J^{(1)}_{ij;t}$ and $ J^{(2)}_{ij;t}$.
In order to take the limit $dt\to 0$, the matrices need
reparametrizing as:
\bea
M_{ij;t}&=&\delta_{ij} + dt\, J_{ij;t}\\
(A_t^{-1})_{ij}&=& dt\, X_{ij;t}.
\eea
Then the random walk reduces to the Langevin equation:
\beq\label{eq1}
\frac{d\vec s_i}{dt}=-\vec s_i \times \left(\vec s_i \times \left(\sum_{j} J_{ij}(t) \vec s_j+\vec\xi_i\right)\right)
\eeq
where $J_{ij}(t)$ denotes the influence of bird $j$ on bird $i$'s
orientation, and $\vec \xi(t)$ is a Gaussian random $d$-dimensional
noise with $\<\vec \xi_i(t) \vec \xi_j(t')\>=dX_{ij}(t)\delta(t-t')$. To
simplify, we assume that $X_{ij}(t)=2T\delta_{ij}$;
$T$ quantifies the noise in alignment, and can be mapped onto a temperature, as we'll see later. In the following, for ease of notation we drop the dependency of $J_{ij}$ on $t$.

The triple cross-product is easier to understand if we note that, for any vector $\vec a$, this cross-product reduces to
\beq
-\vec s \times (\vec s\times \vec a)= \vec a - (\vec s \cdot \vec a)\vec s\equiv \vec a_{\perp},
\eeq
which is just the projection of $\vec a$ onto the hyperplane orthogonal to $\vec s$. Since $\vec s_i$ lives on the unit sphere, its variations must be perpendicular to itself. The triple cross-product just implements this projection by subtracting the parallel part.
This projection ensures the conservation of the norm:
\beq
\frac{d\Vert \vec s_i\Vert^2}{dt}=2\vec s_i \cdot \frac{d\vec s_i}{dt}=0.
\eeq
The norm of $\vec s_i$ stays constant and equal to one.

We rewrite $J_{ij}=Jn_{ij}$, where $J$ quantifies the aligning strength, and $n_{ij}$ how $j$ is taken into account by $i$ ($n_{ij}$ does not have to be an integer). $J$ has the dimension of an inverse time, $n_{ij}$ is dimensionless.
Since anything inside the parentheses of Eq.~\ref{eq1} that is parallel to $\vec s_i$ is discarded, we can rewrite it as:
\beq\label{eq4}
\frac{d\vec s_i}{dt}=J\vec s_i \times \left(\vec s_i \times \left(\sum_{j} \Lambda_{ij} \vec s_j\right)\right)+\vec\xi_{i\perp}
\eeq
where we have denoted $\Lambda_{ij} = \sum_{k} n_{ik} \delta_{ij} -
n_{ij}$, and where now $\<\vec \xi_{i\perp}(t) \vec
\xi_{j\perp}(t')\>=2(d-1)T\delta_{ij}\delta(t-t')$. The $(d-1)$ factor
replaces $d$ because of the projection of the noise term onto the
hyperplane orthogonal to $\vec s_i$. The diagonal term in
$\Lambda_{ij}$ was chosen so as to balance each row of the matrix
($\sum_j \Lambda_{ij}=0$).

There is a link with the statistical description of flock configurations inferred in \cite{Bialek2012}. If $\Lambda_{ij}$ is symmetric and constant in time, the steady-state probability distribution of the set of $(\vec s_1,\ldots,\vec s_N)$ is given by the Boltzmann distribution
\beq\label{eq5}
P(\vec s_1,\ldots,\vec s_N) \propto \exp\left[-\frac{H(s)}{T}\right]
\eeq
with Hamiltonian:
\beq\label{eq6}
H(s)=-\frac{J}{2}\sum_{ij} n_{ij} \vec s_i \vec s_j.
\eeq

We can expand Eq.~\ref{eq4} within the spin-wave approximation. In this limit, all vectors $\vec s_i$ almost point in a common direction, denoted by $\vec n$, so that we can write $\vec s_i = \vec \pi_i + \sqrt{1-\vec \pi_i^2} \vec n \approx  \vec \pi_i + (1-\vec \pi_i^2/2) \vec n$, where $\vec \pi_i$ is the projection of $\vec s_i$ onto the hyperplane orthogonal to $\vec n$: $\vec n\cdot \vec \pi_i=0$. Expanding at first order yields:
\beq\label{eq7}
\frac{d\vec \pi_i}{dt}=- J \sum_{j} \Lambda_{ij} \vec \pi_j+\vec\xi_{i\perp}.
\eeq
In practice, this is the equation we will use for the inference.

\begin{table}
\begin{tabular}{l c c c c c}
Event ID & $N$ & $T$ (s) & $P$ & $v_0$ (m/s) & $r_0$ (m)\\
\hline\hline
20110208\_ACQ3 & 179 & 5.5 & 0.984 & 8.7 & 0.85 \\
\hline
20110211\_ACQ1 & 595 & 4.5 & 0.971 & 8.5 & 0.95 \\
\hline
20110217\_ACQ2 & 407 & 2.1 & 0.986 & 11.0 & 0.70 \\
\hline
20111124\_ACQ1 & 125 & 1.8 & 0.993 & 11.1 & 0.66 \\
\hline
20111125\_ACQ1 & 50 & 5.6 & 0.987 & 12.4 & 1.21 \\
\hline
20111125\_ACQ2 & 530 & 4.4 & 0.957 & 9.2 & 0.85 \\
\hline
20111201\_ACQ3\_1 & 137 & 2.9 & 0.987 & 10.1 & 0.74 \\
\hline
20111201\_ACQ3\_4 & 489 & 2.3 & 0.9763 & 10.5 & 0.74 \\
\hline
20111214\_ACQ4\_1 & 157 & 2.9 & 0.993 & 11.4 & 0.74 \\
\hline
20111214\_ACQ4\_2 & 162 & 4.1 & 0.973 & 11.6 & 1.08 \\
\hline
20111215\_ACQ1 & 401 & 5.7 & 0.987 & 11.0 & 0.82 \\
\hline 
20111220\_ACQ2 & 200 & 1.7 & 0.984 & 16.2 & 0.62 \\
\hline 
20111222\_ACQ1 & 59 & 3.5 & 0.984 & 11.7 & 1.24 \\
\hline
20120209\_ACQ1 & 412 & 3.5 & 0.997 & 29.2 & 0.80 \\
\end{tabular}
\caption{Summary of the data used in the analysis. $N$ is the number of birds, $T$ the duration of the film, $P=(1/N)\Vert\sum_i \vec s_i\Vert$ the polarization of the flock, $v_0$ the average bird velocity, and $r_0$ the average interbird distance. The event ID contains its date and its acquisition index.}
\end{table}

\begin{figure}
\includegraphics[width=.99\linewidth]{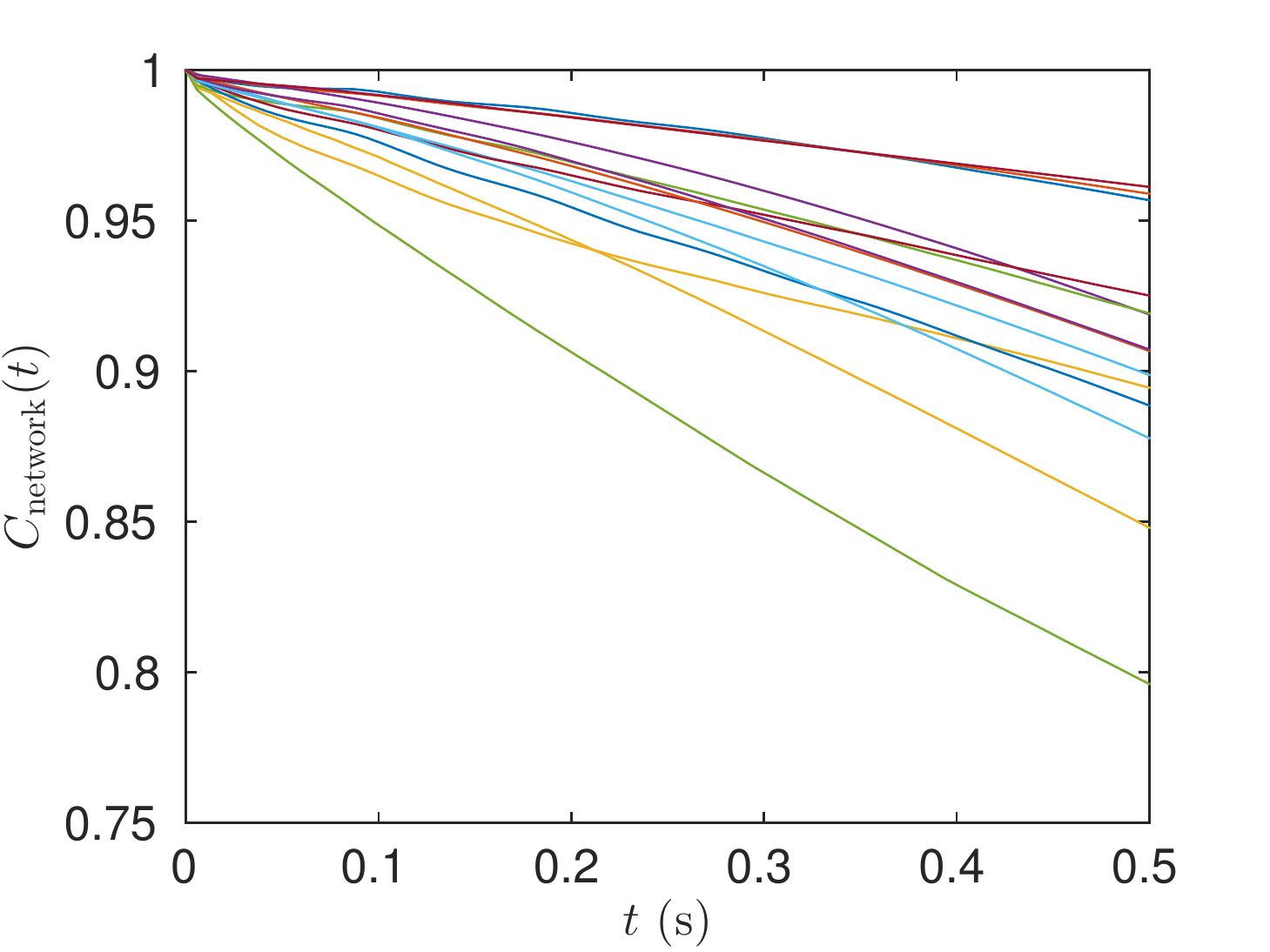}
\caption{Normalized autocorrelation function of the network for all 14 flocking events. The decay is approximately exponential, allowing for the definition of a characteristic decay time $\tau_{\rm relax}$ for each event.}
\label{overlap}
\end{figure}

\begin{figure*}
{
\noindent\includegraphics[width=.329\linewidth]{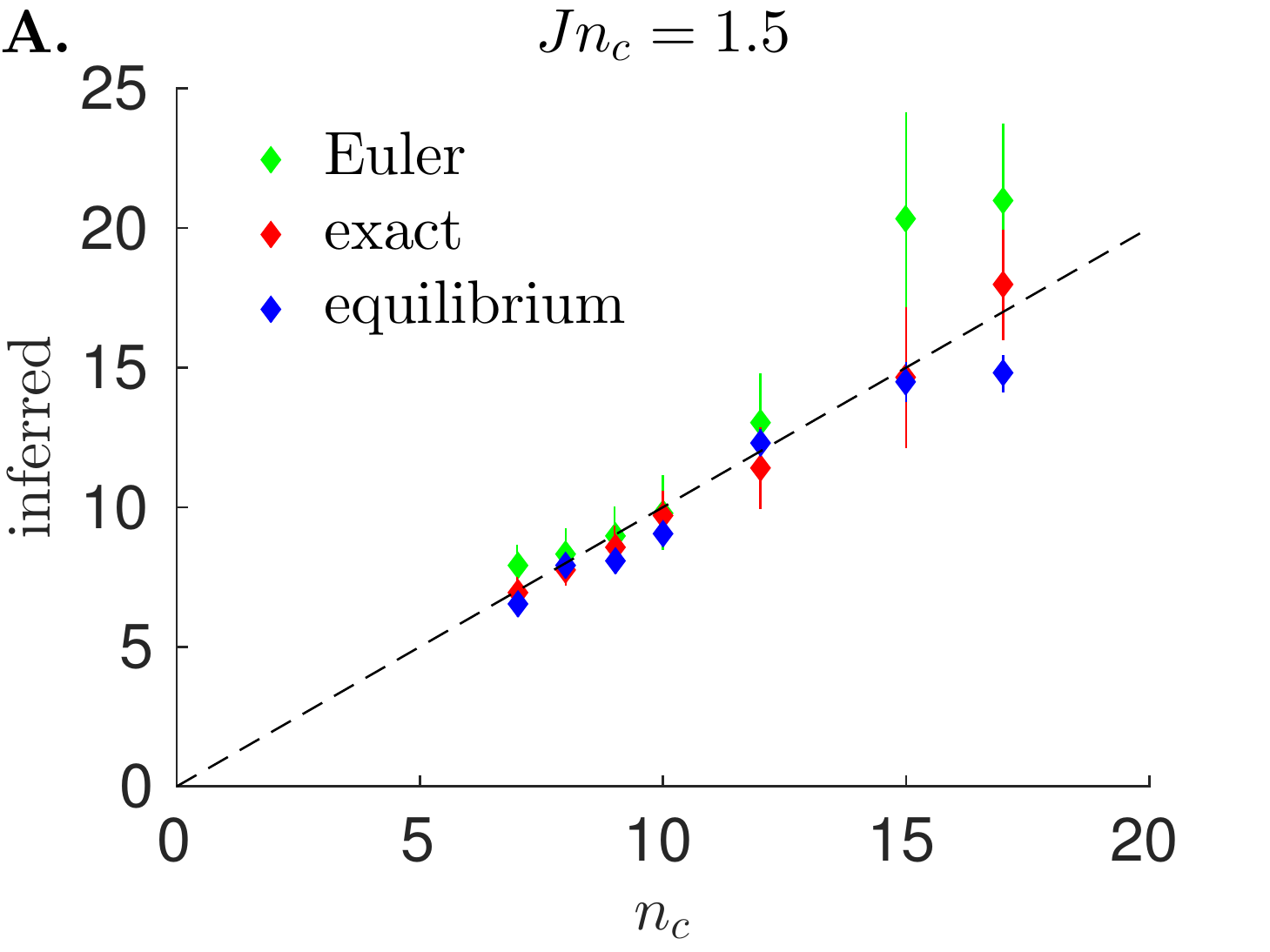}
\includegraphics[width=.329\linewidth]{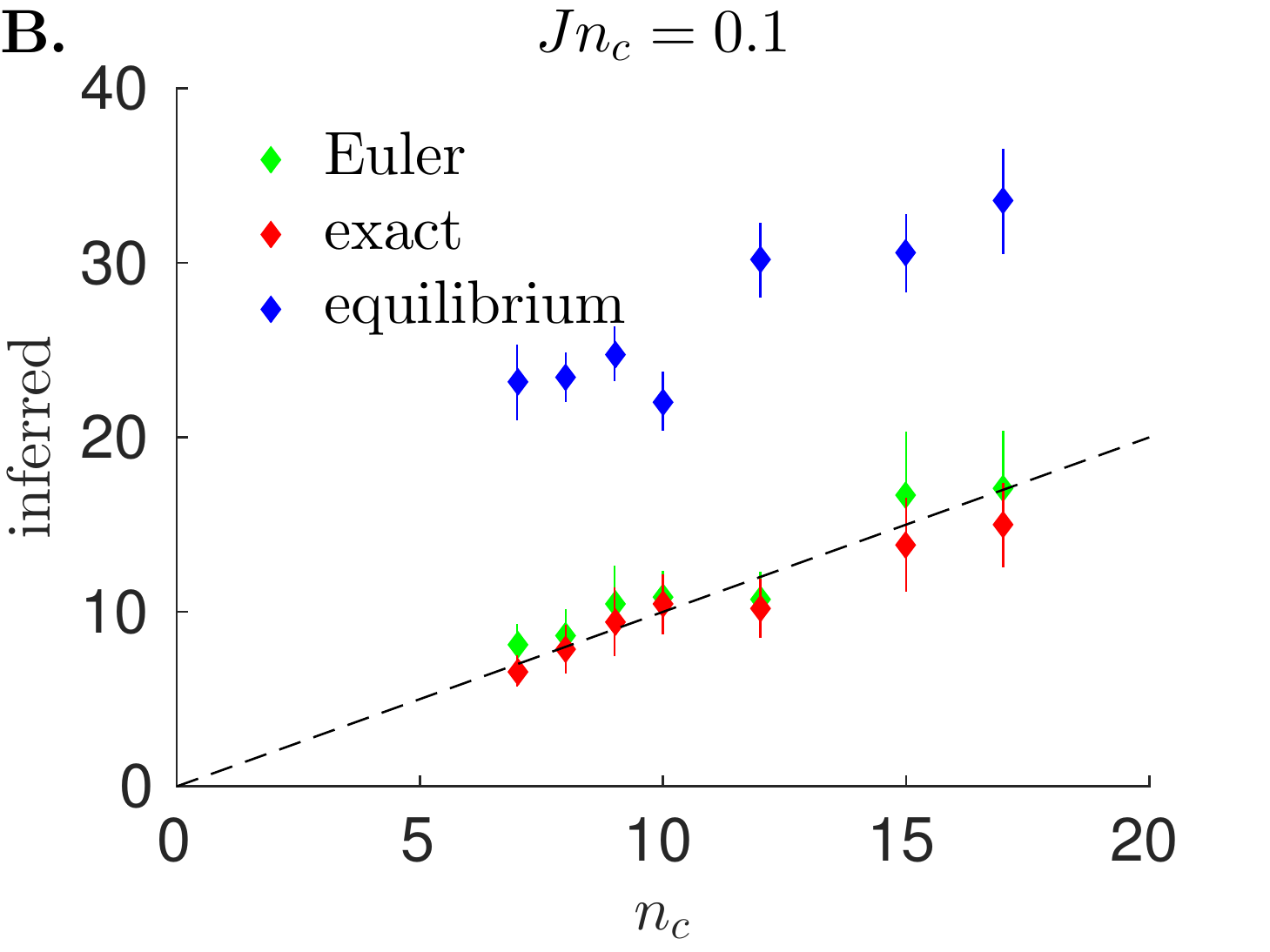}
\includegraphics[width=.329\linewidth]{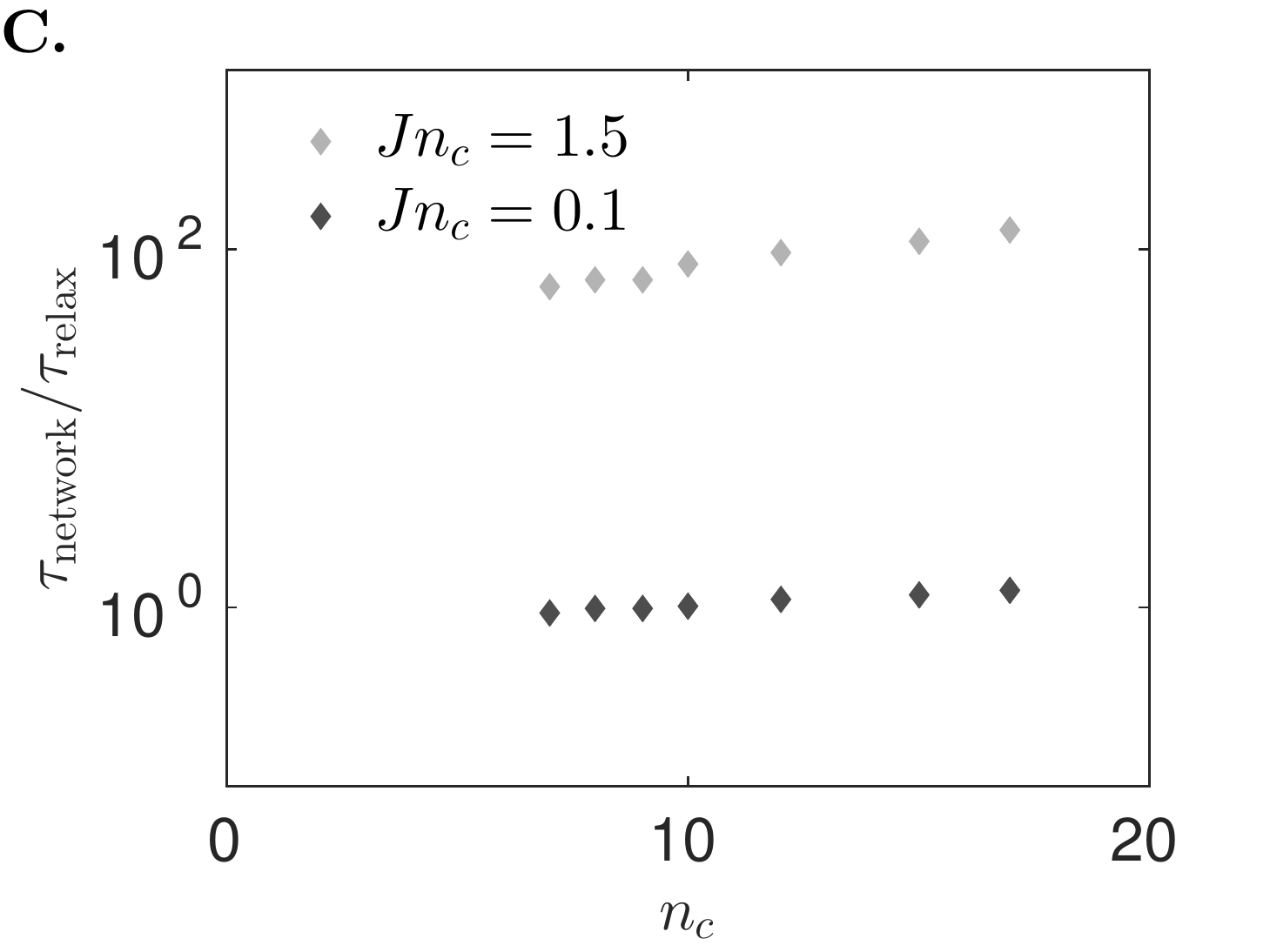}
\caption{Simulations of fast versus slow relaxation.
{\bf A.} Inferred interaction range $n_c$ using dynamical Euler (green), dynamical exact integration (red), or equilibrium-like inference (blue), versus the true $n_c$ for fast relaxation dynamics relative to network rearrangement. The parameters are: $Jn_c=1.5$, $\sqrt{2T}=0.15$, bird speed $v_0=1$, unit bird density (512 birds in an 8 x 8 x 8 box with periodic boundary conditions), inference $dt=0.2$. Polarization is $\approx 0.99$. The equilibrium inference gives the same result as the dynamical one, since the orientation dynamics is fast compared to network reshuffling.
{\bf B.} Same as A., but with slow relaxation of orientations. The parameters are chosen to keep a similar polarization of $0.99$: $Jn_c=0.1$, $\sqrt{2T}=0.05$, bird speed $v_0=1$, unit bird density, inference $dt=1$. The equilibrium inference systematically overestimates the true $n_c$, while the dynamical inferences predict it accurately.
{\bf C.} Comparison of $\tau_{\rm network}$ and $\tau_{\rm relax}$ in the two simulations of {A.} and {B.} The relaxation time $\tau_{\rm relax}$ is taken to be $1/(Jn_c)$, while $\tau_{\rm network}$ is estimated as explained in the main text, by fitting an exponential decay to the overlap autocorrelation function, as in Fig.~\ref{overlap}.
\label{testtimescales}
}
}
\end{figure*}

\begin{figure*}
\includegraphics[width=.49\linewidth]{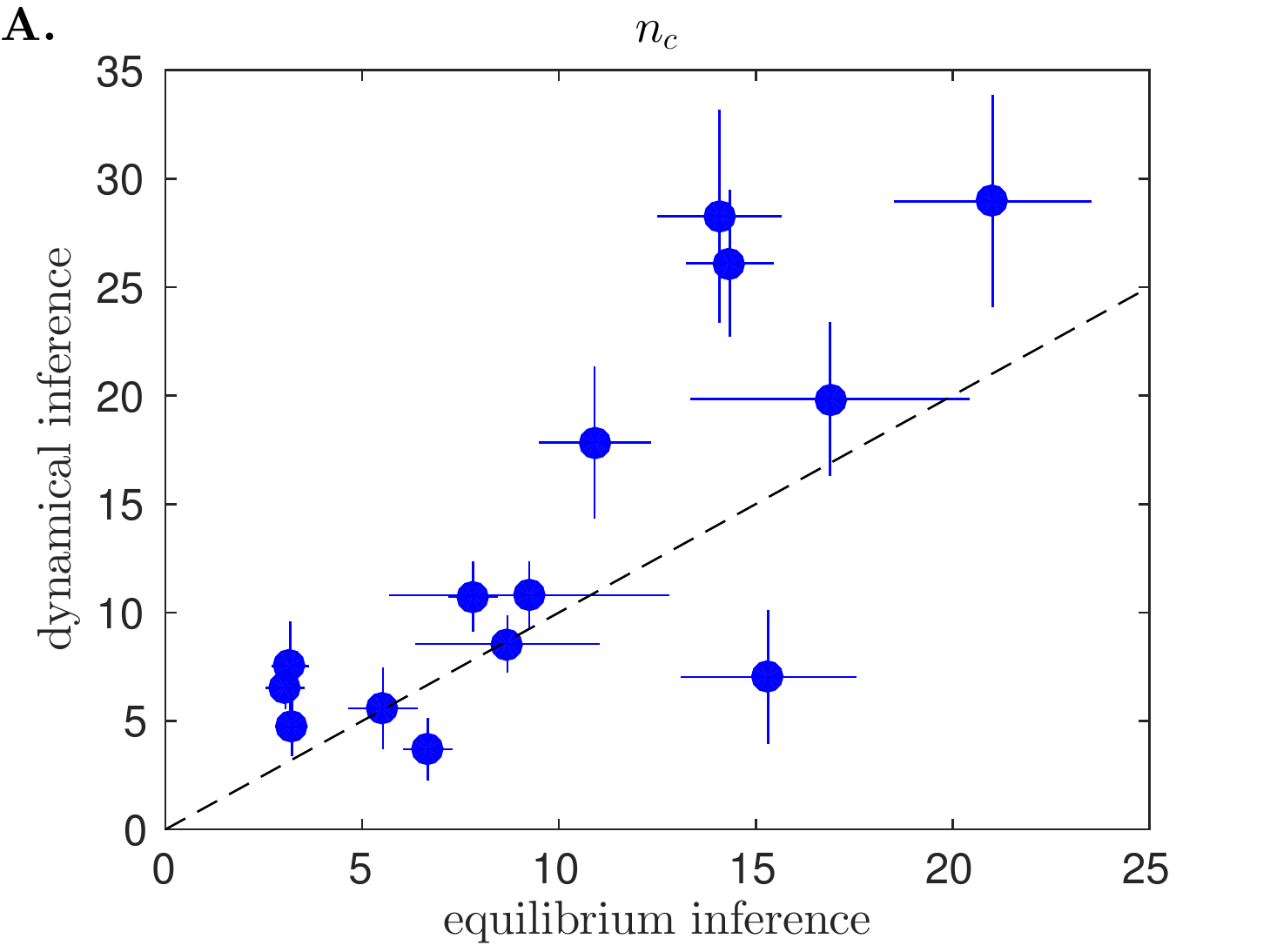}
\includegraphics[width=.49\linewidth]{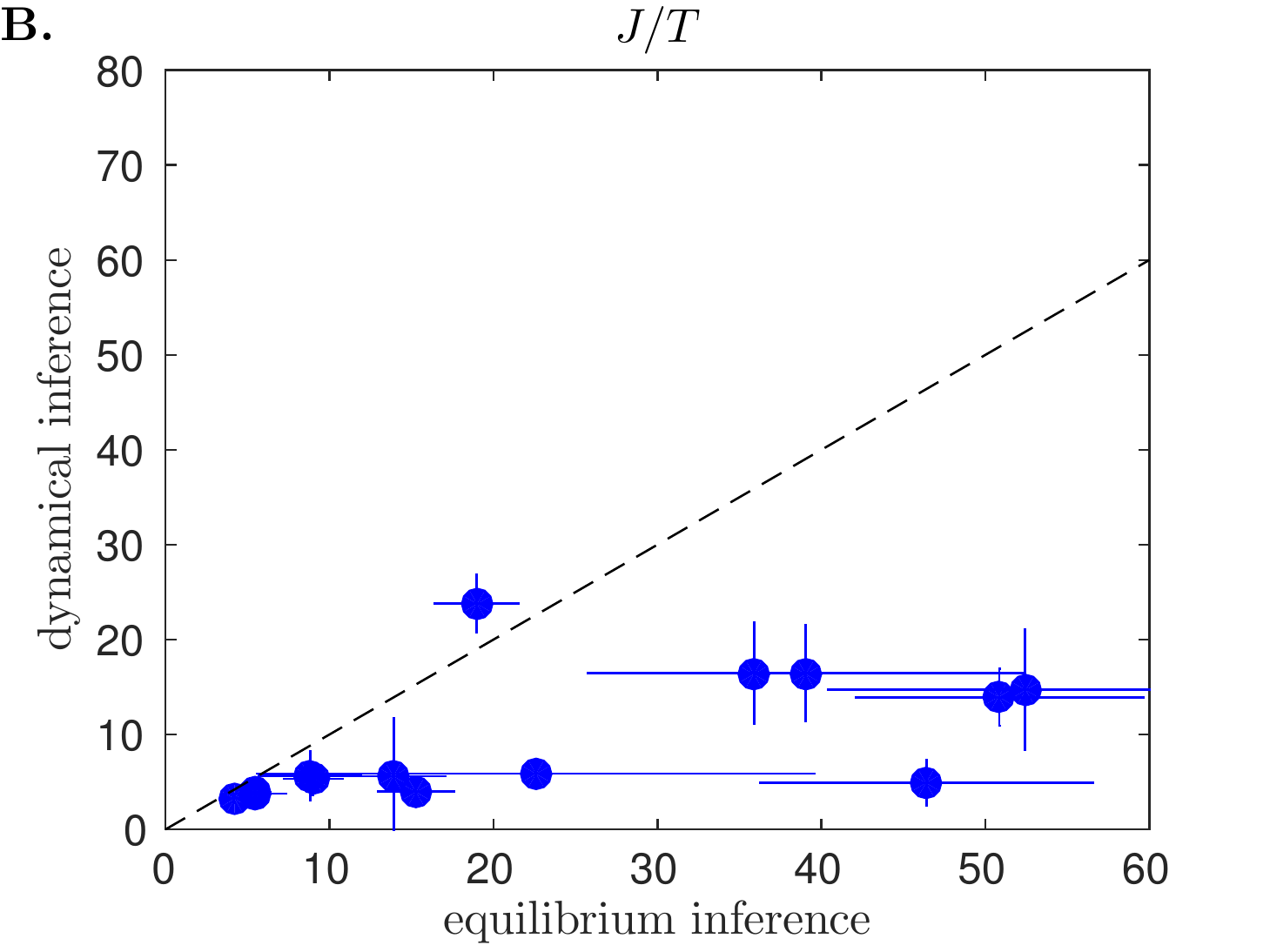}
\caption{Comparison between the equilibrium inference method (abcissa) and
  the dynamical inference method using Euler's rule (ordinate), for
  ({\bf A}) the interaction range $n_c$ and ({\bf B}) the interaction
  parameter $J/T$.
The agreement is relatively poor, especially for the prediction of $J/T$.
}
\label{eulersucks}
\end{figure*}

\begin{figure*}
\includegraphics[width=.49\linewidth]{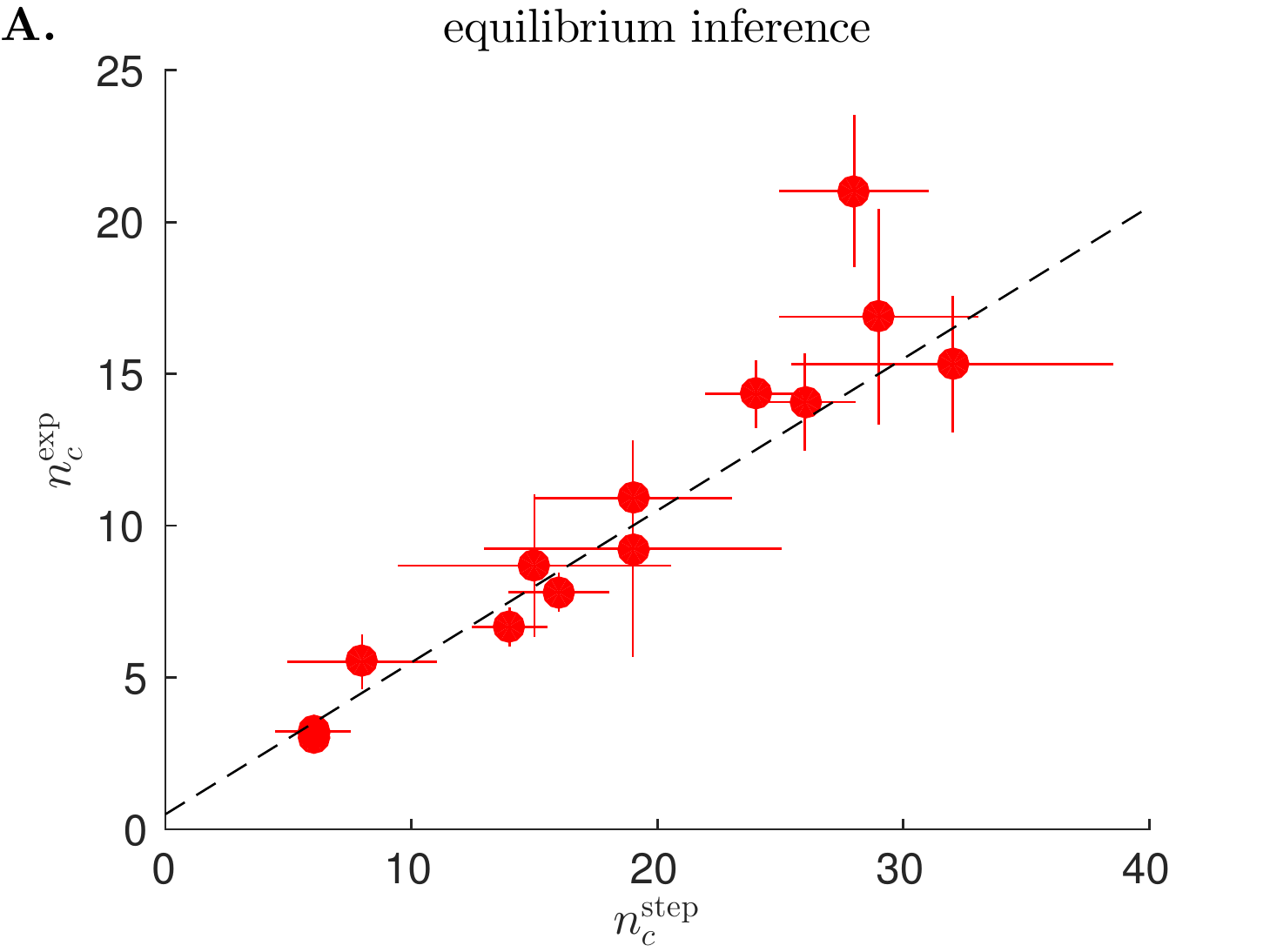}
\includegraphics[width=.49\linewidth]{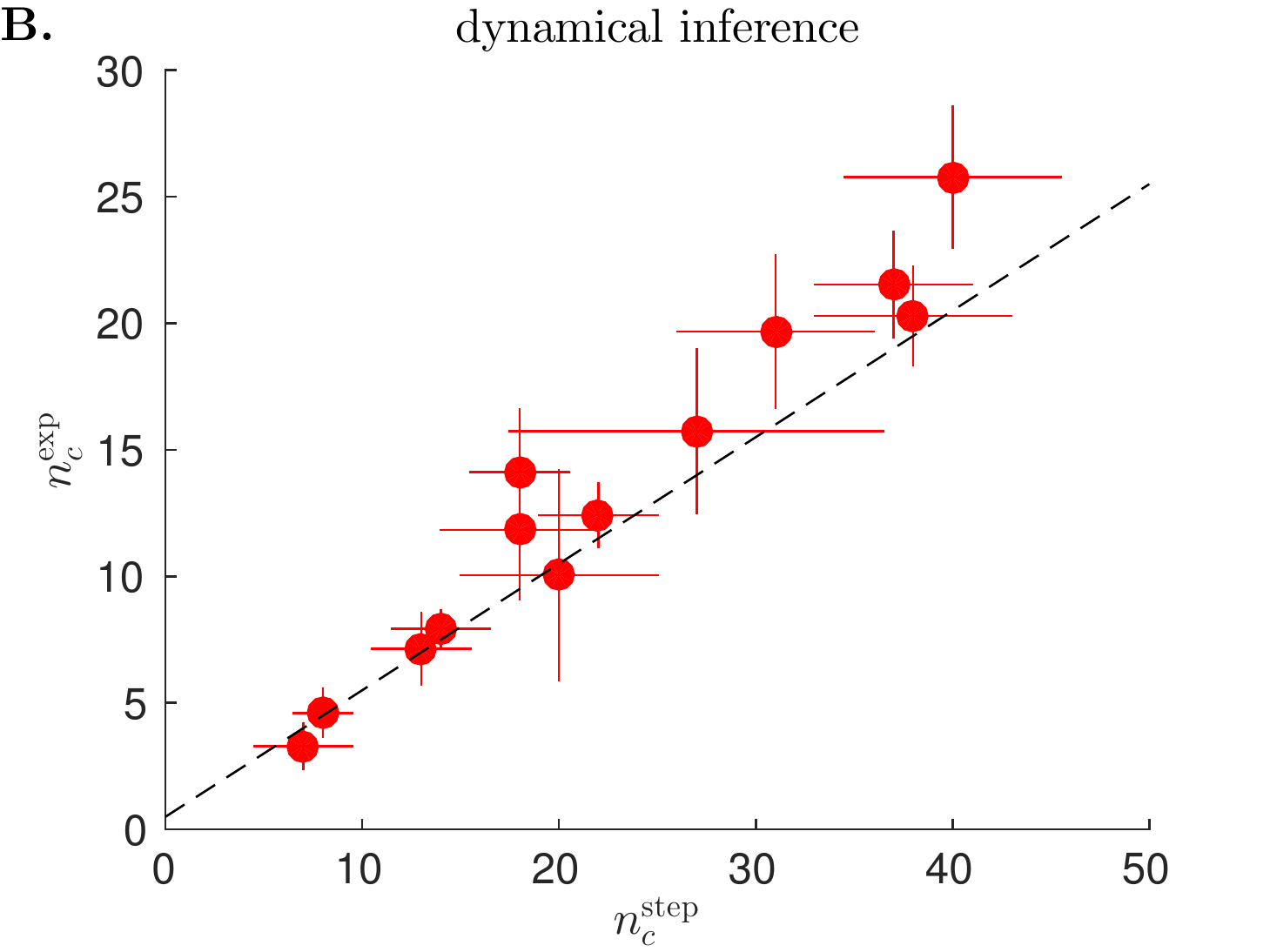}
\caption{Comparison of the interaction range $n_c$ inferred
  assuming a step-function interaction function ($n_c^{\rm step}$, abscissa) or an
  exponentially decaying interaction function ($n_c^{\rm exp}$,
  ordinate), using ({\bf A}) the equilibrium inference method and
  ({\bf B}) the dynamical inference method.
We expect a correspondance between $n_c^{\rm step}$ and
  $n_c^{\rm exp}$: $n_c^{\rm exp} =n_c^{\rm step}/2$. Here
  this correspondance is verified for both inference methods.}
\label{step}
\end{figure*}

\section{Inference from data}
\subsection{Static inference}
We start by recalling how to do the steady-state inference based on the steady-state distribution of Eqs.~\ref{eq5} and \ref{eq6}. We assume that the flock is very polarized, so that the spin-wave approximation is valid. In this approximation, the steady-state distribution reads:
\beq
P(\vec {\bm \pi}|\vec n)
=\frac{1}{Z} \exp\left(-\frac{J}{2T} \sum_{ij} \Lambda_{ij}\vec \pi_i\vec\pi_j\right)\delta\left(\sum_i \vec \pi_i\right)
\eeq
where the common direction $\vec n$ is chosen so that $\sum_i \vec \pi_i=\vec 0$, and where for simplicity $n_{ij}$ is assumed to be symmetric. Integrating over $\vec{\bm \pi}$ satisfying that condition gives the normalization constant:
\beq
Z={\left(\frac{2\pi T}{J}\right)}^{(N-1)(d-1)/2}\prod_{\lambda_k>0} \lambda_k^{-(d-1)/2}
\eeq
where $\lambda_k$ are the eigenvalues of the matrix $\Lambda_{ij}$. Since $\sum_{j} \Lambda_{ij}=0$ for all $i$, we know that one of these eigenvalues is $0$. It is the one corresponding to variations along the direction $(1,\ldots,1)$. These variations are entirely suppressed by the condition $\sum_i \vec\pi_i=0$, and this direction does not contribute to the Gaussian integral, hence the condition $\lambda_k>0$.

In summary, the minus-log-likelihood of the data reads:
\begin{widetext}
\beq\label{eq12}
-\ln P(\vec{\bm \pi}|\vec n)=\frac{J}{2T} \mathrm{Tr}({\bf C}{\bm \Lambda}^{\dagger}) - \frac{(d-1)(N-1)}{2}\ln\left(\frac{J}{T}\frac{1}{2\pi}\right) -\frac{d-1}{2}\sum_{\lambda_k>0}\ln \lambda_k,
\eeq 
where ${\bf C}=\vec {\bm \pi}\vec{\bm  \pi}^\dagger$.

We want to minimize this quantity according to the principle of maximum likelihood. Taking the derivative with respect to $J/T$ gives:
\beq
(J/T)^* = \frac{(d-1)(N-1)}{\mathrm{Tr}({\bf C}{\bm \Lambda}^{\dagger})}\approx \frac{d-1}{C_{\rm int}}
\eeq
with the definition $C_{\rm int}=(1/N)\mathrm{Tr}({\bf C}{\bm \Lambda}^{\dagger})$.

Replacing into Eq.~\ref{eq12} gives:
\beq
-\ln P(\vec{\bm \pi}|\vec n,(J/T)^*)=\frac{(d-1)(N-1)}{2}\left[1+\ln C_{\rm int} +\ln(2\pi/(d-1))\right] - \frac{d-1}{2} \sum_{\lambda_k>0}\ln \lambda_k.
\eeq
\end{widetext}
Finally, this quantity must be minimized over the parameters defining
$\Lambda_{ij}$, or equivalently, ignoring the constants and prefactors:
\beq
\ln C_{\rm int} - \frac{1}{N-1}\sum_{\lambda_k>0}\ln \lambda_k.
\eeq

\subsection{Dynamical inference using Euler's method}
We now move to the dynamical inference from data using Eq.~\ref{eq7}.
Let us start by assuming that we have a series of data points separated by a small $dt$. We can write Euler's approximation to the stochastic differential equation:
\beq\label{euler}
\vec \pi_i(t+dt)=\vec\pi_i(t)- J dt \sum_{j} \Lambda_{ij} \vec \pi_j+\vec\epsilon_{i}
\eeq
where $\vec \epsilon_i$ is Gaussian noise of variance $2(d-1)Tdt$.

Or, in matrix form:
\beq
\vec {\bm \pi}(t+dt)=\vec{\bm\pi}(t)- J dt {\bm\Lambda} \vec {\bm \pi}+\vec{\bm\epsilon}.
\eeq
Let us denote $\vec{\bm\pi}'=\vec {\bm \pi}(t+dt)$. Then the probability of $\vec {\bm \pi}'$ given $\vec{\bm\pi}$ is:
\begin{widetext}
\beq
P(\vec {\bm \pi}'|\vec {\bm \pi})=
 {(4\pi Tdt)}^{-N(d-1)/2}\exp\left[-\frac{1}{4Tdt}(\vec {\bm \pi}'-\vec {\bm \pi}+Jdt {\bm\Lambda} \vec {\bm \pi})^2\right].
\eeq
The associated minus-log-likelihood, $\mathcal{L}=-\ln P(\vec {\bm \pi}'|\vec {\bm \pi})$, is thus given by:
\beq
\mathcal{L}=N\frac{d-1}{2}\ln(4\pi Tdt) + \frac{1}{4Tdt} \mathrm{Tr}\left[{\bf C}
'+{\bf C} -2 {\bf G}+ 2Jdt ({\bf G}-{\bf C}){\bm \Lambda}^\dagger +(Jdt)^2 {\bm \Lambda} {\bf C}{\bm \Lambda}^{\dagger}\right],
\eeq
where ${\bf C}=\vec {\bm \pi}\vec{\bm  \pi}^\dagger$, ${\bf C}'=\vec {\bm \pi}'\vec{\bm  \pi}'^\dagger$ and ${\bf G}=\vec {\bm \pi}'\vec{\bm  \pi}^\dagger$. Or, in short-hand:
\begin{align}
\frac{\mathcal{L}}{N}&=\frac{d-1}{2}\ln(4\pi Tdt) + \frac{1}{4Tdt} \left[C'_{\rm s}+C_{\rm s} -2 G_s+ 2Jdt (G_{\rm int}-C_{\rm int}) +(Jdt)^2 C_{\rm int^2}\right]\\
&\equiv\frac{d-1}{2}\ln(4\pi Tdt) + \frac{\hat{\mathcal{L}}}{4Tdt},
\end{align}
\end{widetext}
with $C'_{\rm s}=\mathrm{Tr}(\mathbf{C'})/N$,
$C_{\rm s}=\mathrm{Tr}(\mathbf{C})/N$, $G_{\rm s}=\mathrm{Tr}(\mathbf{G})/N$,
$G_{\rm int}=\mathrm{Tr}(\mathbf{G}{\bm \Lambda}^\dagger)/N$, $C_{\rm
  int}=\mathrm{Tr}(\mathbf{C}{\bm \Lambda}^\dagger)/N$, and $C_{\rm int^2}=\mathrm{Tr}({\bm \Lambda}\mathbf{G}{\bm \Lambda}^\dagger)/N$

Following the principle of maximum likelihood, which is equivalent to solving the inverse maximum entropy model in the spin-wave approximation, we minimize this quantity over the parameters $J,T$, and the parameters of $\Lambda_{ij}$. Let us start with the temperature $T$. $\partial \mathcal{L}/\partial T=0$ gives:
\beq
T^*=\frac{\hat{\mathcal{L}}}{2(d-1)dt}.
\eeq
We can now minimize $\mathcal{L}$ taken at that value of $T=T^*$,
\beq
\frac{\mathcal{L}(T^*)}{N}=\frac{d-1}{2}\left[1+\ln\hat{\mathcal{L}}+\ln(2\pi/(d-1))\right].
\eeq
In other words, we want to minimize $\hat{\mathcal{L}}$ over the remaining parameters $J$ and $n_c$. Writing the condition for $J$, $\partial \hat{\mathcal{L}}/\partial J=0$ gives:
\beq
J^*=\frac{C_{\rm int}-G_{\rm int}}{dt C_{\rm int^2}}.
\eeq
And replacing into $\hat{\mathcal{L}}$ gives:
\beq
\hat{\mathcal{L}}(J^*)=C'_{\rm s}+C_{\rm s} -2 G_s - \frac{(G_{\rm int}-C_{\rm int})^2}{C_{\rm int^2}}.
\eeq
The first three terms do not depend on the choice of $\Lambda$. The last step is to maximize ${(G_{\rm int}-C_{\rm int})^2}/{C_{\rm int^2}}$ over the paramters defining $\Lambda_{ij}$.

\subsection{Dynamical inference using exact integration}
In general $n_{ij}$ and $\Lambda_{ij}$ may depend on time, because they will evolve with the local neighbours of each birds. But on short time scales such that neighbours do not change significantly, we can view them as constant. If on this time scale the main direction of the flock has not changed much, we can consider Eq.~\ref{eq7} as valid with constant $\Lambda_{ij}$. This linear stochastic equation can actually be solved analytically:
\beq
\vec{\bm \pi}(t+dt)=e^{-J{\bm \Lambda}dt}\vec{\bm \pi}(t) + \int_0^{dt} du\,e^{-J{\bm \Lambda}(dt-u)}\vec{\bm \xi}_\perp(t+u).
\eeq
We define the integrated noise term as:
\beq
\vec{\bm \epsilon}=\int_0^{dt} du\,e^{-J{\bm \Lambda}(dt-u)}\vec{\bm \xi}_\perp(t+u).
\eeq
Since it is a sum of Gaussian variables, $\vec{\bm\epsilon}$ is also Gaussian, of mean zero and covariance:
\beq
\<\vec{\bm \epsilon} \vec{\bm \epsilon}^{\dagger}\>= 2(d-1)T \int_0^{dt} du\,e^{-J{\bm \Lambda}u}e^{-J{\bm \Lambda}^\dagger u}
\eeq
In the limit $dt\to 0$, we recover Euler's approximation, Eq.~\ref{euler}.

With this new, exact integration formula, we can write the minus-log-likelihood:
\beq
\mathcal{L}=N\frac{d-1}{2}\ln(4\pi Tdt) + \frac{d-1}{2}\ln\det {\bf B}+N\frac{\hat{\mathcal{L}}}{4Tdt},
\eeq
with:
\beq
\hat{\mathcal{L}}=\frac{1}{N}\mathrm{Tr}\left[{\bf C}'{\bf A} -2{\bf G}e^{-J{\bm\Lambda}^\dagger dt}{\bf A}+e^{-J{\bm\Lambda} dt}{\bf C}e^{-J{\bm\Lambda}^\dagger dt}\right],
\eeq
\beq
{\bf A}={\bf B}^{-1}\quad\textrm{and}\quad {\bf B}=\frac{1}{dt}\int_0^{dt} du\,e^{-J{\bm \Lambda}u}e^{-J{\bm \Lambda}^\dagger u}.
\eeq

As before, we can solve for $T$ easily:
\beq
T^*=\frac{\hat{\mathcal{L}}}{2(d-1)dt},
\eeq
yielding:
\beq
\frac{\mathcal{L}(T^*)}{N}=\frac{d-1}{2}\left[1+\ln\hat{\mathcal{L}}+\frac{1}{N}\ln\det {\bf B}+\ln(2\pi/(d-1))\right].
\eeq
 Note that now ${\bf A}$ and therefore ${\bf B}$ depend on $J$ as well as $\Lambda_{ij}$. The sum $[\ln\hat{\mathcal{L}}+(1/N)\ln\det {\bf B}]$ must be minimized  numerically with respect to both $J$ and the parameters defining ${\bm \Lambda}$.

\subsection{Two parametrizations for $n_{ij}$}

We now need to specify the matrix $\Lambda_{ij}$. Here we only consider topological distance for the interaction matrix. Let us denote $k_{ij}$ the rank of $j$ among the neighbors of $i$, from the closest in distance to the farthest.

In the first parametrization, already used in previous work, we say that a bird interacts with its $n_c^{\rm step}$ closest neighbours. This corresponds to:
\begin{equation}
\textrm{step:}\qquad n_{ij}=\Theta(n_c^{\rm step}-k_{ij}),
\end{equation}
where $\Theta(x)=1$ if $x\geq 0$ and $0$ otherwise.
Numerically, $J^*$ is calculated for each integer value of $n_c^{\rm step}$ using a simple iterative 1D optimization algorithm.

In the second parametrization, we assume an exponentially decaying interaction as a function of rank:
\beq
\textrm{exp:}\qquad n_{ij}=\exp(-k_{ij}/n_c^{\rm exp}).
\eeq
Numerically, we implement a 1D iterative optimization algorithm for $n_c^{\rm step}$, where $J^*(n_c^{\rm exp})$ is calculated for each $n_c^{\rm step}$ as before, in a nested loop.

Can we compare the two parametrizations?
In the first case, the average rank of an interacting neighbour is $(n_c^{\rm step}+1)/2 \approx n_c^{\rm step}/2$. In the second case, this average rank is $\approx n_c^{\rm exp}$.
It makes sense to hypothesize this average rank should be invariant, regardless of the choice of parametrization. Then, if we infer models with data using the two parametrizations, we expect:
\beq
n_c^{\rm exp}\approx \frac{n_c^{\rm step}}{2}.
\eeq

The second important effective parameter is the total interaction strength $J\sum_j n_{ij}$, equal to $J_{\rm step}n_c^{\rm step}$ is the first case, and to $\approx J_{\rm exp}n_c^{\rm exp}$ in the second one. Requiring that these quantities are equal in the two parametrizations yields:
\beq
J_{\rm exp}\approx 2 J_{\rm step}.
\eeq

Figure \ref{step} shows that the effective $n_c^{\rm step}$ and $n_c^{\rm exp}$ learned from data follow these relations accurately.

\section{Orientation relaxation time}

In our work we compare the relaxation time of the orientational degrees of freedom, $\tau_\mathrm{relax}$, to the reshuffling time of the network, $\tau_\mathrm{network}$, finding the first one to be much smaller than the second one. This may seem an odd result, as in a fixed-lattice theory with spontaneously broken continuous symmetry both the correlation length and the relaxation time {\it diverge} with the system size $L$. Hence, in what sense can $\tau_\mathrm{relax}$ be small?

In the following we consider a fixed
lattice for the following reason: we need to compare the relaxation time
to the network reshuffling time; to do this consistently, we need to work out the 
relaxation time of the order parameter {it in absence} of the effect of 
network reshuffling.
To fix ideas we also work on a regular lattice in the continuum limit; the
following arguments, though, are valid in general.
 In this limit Eq.~\ref{eq7} now reads:
\beq
\frac{d\vec\pi}{dt}=Jn_ca^2 \Delta \vec\pi + \vec\xi_\perp.
\eeq
where $\Delta$ is the Laplacian operator and $a$ the lattice spacing. In Fourier space, this
equation becomes:
\beq
i\omega \vec\pi(k,\omega) = -{Jn_c (ka)^2} \vec\pi(k,\omega) + \vec\xi_\perp(k,\omega)
\eeq
and its solution is:
\beq
\vec \pi(k,\omega) = G(k,\omega) \vec \xi_\perp(k,\omega),
\label{zuzu}
\eeq
were the dynamical propagator (or dynamic response) of the Gaussian
spin-wave theory in Fourier space is:
\beq
G(k,\omega) = \frac{1}{i\omega + J a^2 n_c k^2 } \ ,
\label{zuzu2}
\eeq
We need now to compute the dynamical self-correlation function, that is the correlation of the fluctuations
at the {\it same} position $x$ (or site $i$), namely,
\beq
C_\mathrm{relax}(t)=\<\vec \pi(x,t_0)\cdot \vec\pi(x,t_0+t)\> \ .
\eeq
From \eqref{zuzu} and \eqref{zuzu2} we have,
\begin{widetext}
\beq
\label{autocor}
C_\mathrm{relax}(t)= 2(d-1)T\int_{1/L}^{1/a} d^dk \ \int d\omega \  \frac{e^{-i\omega t}}{(i\omega + J a^2 n_c k^2 )(i\omega-J a^2 n_c k^2)} = 2(d-1)T
\int_{1/L}^{1/a} d^dk \ \frac{e^{-J a^2 n_c k^2 t}}{J a^2 n_c k^2}  \ ,
\eeq
\end{widetext}
which (up to constant prefactors) is the self-correlation function reported in the main text.
The absence of a mass term (zero mode) implies that in $d=3$ the 
function $C_\mathrm{relax}(t)$ is a power law, so that the self-relaxation time 
diverges with $L$. However, as we explain the main, the modes that contribute 
to the rearrangement of the network are only those with short wavelength, comparable with the interaction 
range $r_c$; hence, only $k$ larger to $1/r_c$ contributes to the network reshuffling in the integral above, and we 
therefore define the effective correlation function, 
\beq
\label{bob}
C_\mathrm{relax}^*(t) \equiv 2(d-1)T \int_{1/r_c}^{1/a} d^dk \  \frac{e^{-J a^2 n_c k^2 t}}{J a^2 n_c k^2} \ .
\eeq
This correlation function has now an exponential behavior for large $t$, with finite relaxation time equal to
$(1/J n_c) \cdot  (r_c/a)^2$. The ratio between interaction range and lattice spacing, $(r_c/a)$, is 
in general of order $1$ for short range interaction (as it is the case in flocks) and therefore the time scale of relaxation of the 
orientational degrees of freedom is $\tau_\mathrm{relax} = (J n_c)^{-1}$, which is what we study in the main text.

If we do not assume a regular lattice, instead of a differential Laplacian operator, we have to deal 
with the generic Laplacian matrix $\bm \Lambda$ in equation \eqref{model2} in the main text, and with 
its eigenvalues, let us call them $\Lambda$. In this case the self-correlation function is given by, 
\beq
\label{bobgen}
C_\mathrm{relax}(t)  \equiv 2(d-1)T   \int_{\Lambda_{min}}^{\Lambda_{max}} d\Lambda \  \rho(\Lambda) \ \frac{e^{-J n_c \, \Lambda t}}{J n_c \, \Lambda} \ ,
\eeq
 where $\rho(\Lambda)$ is the eigenvalue spectrum of $\bm \Lambda$. In a spatially homogeneous network 
 $\Lambda$ scales as an inverse length squared, playing the same role as $k^2$ in a regular lattice. 
 Thus, $\Lambda_{min}\sim 1/L^2$ and $\Lambda_{max}\sim 1/a^2$, $a$ being the average nearest neighbors distance.  
 The absence of a $\Lambda$-independent term at the denominator is equivalent to the absence of a $k$-independent term 
 the case of a regular lattice (zero mode). Similarly, the largest contribution to the integral comes from the 
modes near the lower extreme of integration, $\Lambda_{min}\sim1/L^2$. The previous argument then requires
to restrict the integral for $\Lambda>1/r_c^2$, hence giving
\beq
\label{bobgen2}
C_\mathrm{relax}^*(t) \equiv 2(d-1)T   \int_{1/r_c^2}^{1/a^2} d\Lambda \ \rho(\Lambda) \ \frac{e^{-J n_c \Lambda t}}{J n_c \Lambda} \ .
\eeq
which (as in the regular lattice case) gives exponential relaxation with  $\tau_\mathrm{relax} = (J n_c)^{-1}$.

Our argument to restrict the $k$ integral in the self-correlation function to short wavelength modes, $k>1/r_c$,  
finds a strong consistency check in the following fact: even the network 
correlation function, $C_{\rm
  network}(t)$, does depend on a local scale, exactly as
$C_\mathrm{relax}^*$ depends on $r_c$. When we ask what is the degree
of reshuffling of the interaction network within a time $t$, we are
effectively asking how much the network changes over a spatial scale
$n_c$. We could, for example, ask what is the time needed to disrupt
the entire network, i.e. the reshuffling over a scale $N$, and this
would give a much larger time, scaling with $N$ (for a computation of
this time and its connetcion to mutual diffusion in space see \cite{Cavagna2013a}). In a similar way, when we integrate in \eqref{autocor} down to $1/L$ we get a time scale which scales with $L$. Hence, when comparing orientation relaxation and network reshuffling we need to fix a scale for both phenomena. Since we are interested here in inferring the interaction rules, the right scale is the scale of interaction, namely $r_c$ or $n_c$. On the other hand, as we discuss in the conclusions of the main text, were we interested in studying (or predicting) the large size behaviour in the long time limit, we should assess the divergence of both time scales with the size, which is the realm of the hydrodynamic theory.
{
 In general, both timescales $\tau_{relax}$ and  $\tau_{\rm network}$ can be defined on a given spatial scale $r$ (or the equivalent topological scale $n$). What we expect is that, as this scale $r$ increases,  $\tau_{relax}$ and $\tau_{\rm network}$ become closer and, at given $r^\star$,  one has $\tau_{relax}(r^\star)\sim \tau_{\rm network}(r^\star)$. This lengthscale $r^\star$ represents a crossover scale above which the motility of individuals becomes relevant and the system behaves in a non-equilibrium way. When $r^\star\gg r_c$ we are in the condition of local equilibrium that we discussed in this paper. Note that an estimate of the crossover length can also be computed using scaling arguments within the hydrodynamic approach, see e.g. \cite{toner2005hydrodynamics}}

\end{document}